# Three-dimensional reconstruction of integrated implosion targets from simulated small-angle pinhole images

SHIJIAN LI,[1] QIANGQIANG WANG,[2] XIAN WEI,[1] ZHURONG CAO,[2,3] AND QING ZHAO [1,4]

[1]*Center for Quantum Technology Research, School of Physics, Beijing Institute of Technology, Beijing 100081, China*
[2]*Laser Fusion Research Center, China Academy of Engineering Physics, Mianyang, Sichuan 621900, China*
[3]*cao33jin@aliyun.com*
[4]*qzhaoyuping@bit.edu.cn*

**Abstract:** To indirectly evaluate the asymmetry of the radiation drive under limited measurement conditions in inertial confinement fusion research, we have proposed an integral method to approximate the three-dimensional self-radiation distribution of the compressed plasma core using only four pinhole images from a single laser entrance hole at a maximum projection angle of 10°. The simultaneous algebraic reconstruction technique (SART) that uses spatial constraints provided by the prior structural information and the central pinhole image is utilized in the simulation. The simulation results showed that the normalized mean square deviation between the original distribution and reconstruction results of the central radiation area of the simulated cavity was 0.4401, and the structural similarity of the cavity radiation distribution was 0.5566. Meanwhile, using more diagnostic holes could achieve better structural similarity and lower reconstruction error. In addition, the results indicated that our new proposed method could reconstruct the distribution of a compressed plasma core in a vacuum hohlraum with high accuracy.

## 1. Introduction

In the research of indirectly driven laser fusion, one of the most important issues is evaluating the asymmetry or nonuniformity of the radiation drive [1]. The driven radiation asymmetry can be indirectly evaluated and analyzed by the asymmetry of the implosion hotspot. To obtain the radiation flux information with high temporal resolution and high two-dimensional (2D) spatial resolution, various imaging techniques have been proposed. A single-channel Kirkpatrick–Baez (KB) microscope combined with a streak camera and a four-channel KB microscope combined with an X-ray framing camera have become the key X-ray imaging diagnostic equipment of laser facilities such as Omega and Laser MegaJoule (LMJ) and have been successfully applied to the National Ignition Facility (NIF)[2-4]. A new technique has been developed in China's laser facilities for space-resolving measurement of X-ray flux from a specific area inside the hohlraum [5,6]. Compressed ultrafast photography (CUP) based on a streak camera can achieve ultrashort temporal resolution [7,8]. Some studies have been conducted on the three-dimensional (3D) simulation of indirectly driven laser fusion [9,10]. However, the direct 3D measurement of implosion targets is still difficult to achieve in limited conditions. In 2015, the first integrated implosion experiments in vacuum hohlraums were conducted at the 100-kJ-level laser facility in China [11]. To evaluate the asymmetry of radiation driving, pinhole cameras and a multichannel KB microscope were used to obtain the high spatial resolution shape information of the hotspots from the polar and equatorial directions. In this experiment, the equatorial hotspot images were obtained through a square diagnostic hole in the middle of the vacuum hohlraum. All images were 2D projections of a 3D compressed plasma core. The reconstructed 3D core based on these 2D projections was significantly affected by the number of projections, the range of the projection angles, and the noise. In the inertial confinement fusion (ICF) experiments, the range of projection angles and the number of projections were limited, which related to the number of diagnostic holes [12].

Unlike pinhole array imaging, some penumbra imaging system-based methods used in ICF experiments have been proposed [13-15]. Various pinhole or micro-lens arrays have also been used in integrated imaging to reconstruct and display 3D objects [16-20], and some good results have been achieved. However, the target area for integrated imaging was the surface and contour of the 3D objects, excluding the interior. Meanwhile, some methods have been developed and utilized to deal with the 3D reconstruction of a laser implosion target based on pinhole images. A heuristic method using three pinhole images taken from three viewing directions [21] and an integral method using six pinhole images located on three coordinate axes [22] gave good results. Nevertheless, there was an issue with the wide angles of the pinhole images. Unless more diagnostic holes are added, it is difficult to increase the projection angles for a vacuum hohlraum, and this may affect the asymmetry of the radiation drive.

The simultaneous algebraic reconstruction techniques (SART) [23] have been widely used in computed tomography imaging. Various methods have been added to SART to accelerate the reconstruction, including using 2-D Texture Mapping Hardware [24], variable step size [25], and fast projection algorithm [26]. SART was also improved by an adaptive method based on the idea of carefully choosing the order of data access and adjusting the relaxation parameters. Meanwhile, different methods were incorporated into the SART for specific situations, such as 'C-SART' using the 2D-Laplacian operator for smooth fields in ionospheric tomography [27], A weighting scheme for objects with high-attenuation feature [28], guided image filtering for the situation of which the prior image was known [29], and total variation (TV) minimization for Positron-Emission Tomography [30]. In this study, a method incorporated spatial constraints was proposed based on the SART. The spatial constraints were provided by the prior structural information and central pinhole image. The space-constrained area was computed before iteration, and the rest was target area. During the calculation, only the target area was updated because of the aforementioned limitations with respect to the projection angle range and the number of projections. Under the assumption of a noise- and scatter-free imaging system, a vacuum hohlraum with two laser entrance holes (LEHs) at the top and bottom was imaged. The center of the vacuum hohlraum was the self-radiation area of the compressed core. Four pinhole images at a single LEH and four pinhole images at diagnostic hole located on middle of hohlraum wall were taken, and the maximum projection angle relative to the axis of the hole was 10°. In addition, during the iteration procedure, we updated one section at a time, instead of one voxel, to reduce the computational cost. The simulation results showed that the proposed method could estimate the self-radiation distribution of the vacuum hohlraum center at small projection angles and limited pinhole images, and the accuracy of the reconstructed 3D plasma core was higher than that of the conventional SART method.

## 2. Method

### 2.1 The diagnostic system

The model parameter setting is based on the physical parameter of the experimental device of 100-kJ-level laser facility in China [11], and vacuum hohlraums were used in the experiments. A total of 48 lasers were simultaneously injected into vacuum hohlraums of 4400 μm in length and 2200 μm in diameter from two LEHs. The LEH was 1200 μm in diameter. The diameter of the capsules was approximately 800 μm. A 400μm square diagnostic hole was set on the middle of the hohlraum wall. The structure is given in Fig. 1.

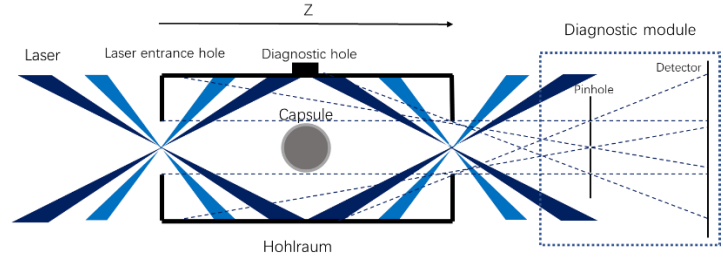

Fig. 1. Schematic of the hohlraum.

The voxel size was set to 20 × 20 × 20 μm³, and the vacuum hohlraum was a matrix of 110 × 110 × 210 voxels, where the Z direction was the axis of the vacuum hohlraum. According to the time-integral imaging of the central plasma core from the polar and equatorial directions of the experiments in Ref. [11], we assumed that the radiation distribution of the core is a sphere that is symmetrical in the X and Y directions, but asymmetrical in the Z direction. As there was a certain radiation distribution on the vacuum hohlraum wall due to re-emission, we set it to a fixed value for simplicity.

The diagnostic module consisted of a pinhole array (in which the diameter of the pinholes was 15 μm) and planar detectors (where the physical size of a pixel was 9 × 9 μm²), as shown in Fig.1. One pinhole was placed at the axis of the vacuum hohlraum in the center of the pinhole array to estimate the size of the central radiation field, and the other three pinholes were evenly distributed within the maximum projection angle. The pinhole array was 100 mm away from the LEH, and the planar detectors were 200 mm away from the pinhole array. With this parameter setting, the magnification factor was about 2. Considering that the minimum laser incident angle was 28°, the maximum projection angle relative to the axis was set to 10° to reduce the impact on the laser injection system. The diameter of the LEH was smaller than that of the hohlraum, which limited the imaging field of view.

## 2.2 Geometric relationship

The geometric relationship between a 3D source hohlraum and its 2D pinhole image is shown in Fig. 2, where O is the section of the hohlraum along the Z direction, P is the pinhole, R is the pinhole image formed on the detector, U is the distance from the LEH to the pinhole array, and V is the distance from the pinhole array to the planar detectors.

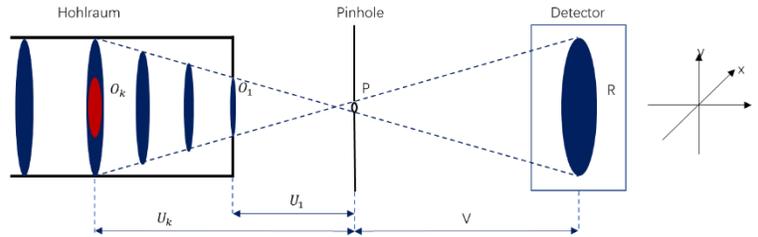

Fig. 2. Geometric relationship of the hohlraum, pinhole, and detector.

According to this geometric relationship, the size R of the image on the planar detectors is determined by:

$$R = \frac{O \cdot U + P \cdot (U + V)}{U}. \tag{1}$$

## 2.3 Conventional SART

The essential idea of the SART algorithm is to give an initial value of the reconstruction area, which is generally zero, and then back-project the residuals of the projection values uniformly along its ray direction. The target image is continuously corrected using the mean of the correction values of all the rays until the projected image is satisfied, and then the iterative procedure ends. The iterative formula of SART is

$$x_j^{k+1} = x_j^k + \lambda \frac{\sum_i (\omega_{ij} \frac{P_i - \sum_{n=1}^{J} \omega_{in} x_n^k}{\sum_{n=1}^{J} \omega_{in}})}{\sum_i \omega_{ij}}, \qquad (2)$$

where $k$ is the number of iterations, $\lambda$ is the relaxation parameter, $P_i$ is the projection value of the $i$-th ray, $\omega_{ij}$ is the weight factor of each voxel, $x_j^k$ is the $j$-th voxel in the $k$-th iteration, and $J$ is the voxel number.

*2.4 SART based on section updating and spatial constraints*

The conventional SART method was updated in units of a single pixel. Considering that the imaging plane of the pinhole array was perpendicular to the axis of the hohlraum, we updated one section at a time instead of one voxel. The above SART iterative formula was updated by

$$X_j^{k+1} = X_j^k + \frac{\lambda}{\sum_i V_{ij}} A^* \{\sum_i [\frac{P_i - \sum_{n=1}^{J} A(X_n^k)}{\sum_{n=1}^{J} W_{in}}]\}, \qquad (3)$$

where $k$ is the $k$-th iteration, $\lambda$ is the relaxation parameter, $A^*$ is the backward projection operator, $A$ is the forward projection operator, $P_i$ is the pinhole image of the $i$-th pinhole, $X_j^k$ is the $j$-th section in the $k$-th iteration, and $J$ is the total number of sections. $W_{in}$ is computed by forward projecting $n$-th section with unit intensity through the $i$-th pinhole and $V_{ij}$ is computed by backward projecting $i$-th pinhole image with unit intensity to the $j$-th section.

For small-angle projections, several possible 3D source distributions satisfy the measured projection data, and it is difficult to find an optimal solution. The iterative solutions found may have a negative value, which is unacceptable for this research. Therefore, all values less than zero were set to zero in the iteration. The spatial constraints were computed before the iteration based on the prior structural information of the hohlraum and the central pinhole image. During the iteration, we maintained the space-constrained area at zero and then performed the SART iteration.

*2.5 Spatial constraints*

Spatial constraints contained two part: the prior structural information (such as diameter and shape) of the hohlraum could extract a spatial mask of the position of hohlraum wall and LEHs. The pinhole image placed at the axis of the hohlraum could form spatial constraints of central radiation area. To leverage these constraints, a spatial mask was formed from the central pinhole image. In this step, the non-zeros area in the middle of pinhole image was first extracted to get a binary mask. The 2D binary mask was then backward projected to middle section of hohlraum as the central self-radiation mask. The central self-radiation area was selected as the sum of the spheroid formed by rotating the central self-radiation mask along the x-axis and y-axis, respectively. From the experimental results in Ref. [11], the central self-radiation area of the hohlraum was essentially a flat ellipsoid, and its diameter at the axial direction was smaller than that at the horizontal direction, which ensured that the estimated central self-radiation area contained the original central self-radiation area of the hohlraum. The space-constrained area was finally selected as the portion from the inside of the hohlraum wall to the outside of the central self-radiation area.

3. **Results and discussion**

To evaluate the accuracy of the proposed method, we simulated a vacuum hohlraum of 110 × 110 × 210 voxels with a simulated compressed plasma core in the center. Based on the experimental results in Ref. [11], two cases were considered: one is that only four pinhole images of a single LEH were used; the other is that four pinhole images of a single LEH and four pinhole images of the diagnostic hole were used in calculation. The maximum projection angle of these pinhole images was 10° to the central axis of the hole. 200 iterations were performed for each calculation. To quantitatively assess the quality of recovery, a normalized mean square error (NMSE) [21] was defined. The NMSE between the source X and the restored Y was defined as:

$$NMSE(X,Y) = \frac{\|X - Y\|^2}{\|X\|^2}, \qquad (4)$$

Meanwhile, to evaluate the similarity of the reconstructed self-radiation distribution, the structural similarity (SSIM) was defined [31] as:

$$SSIM(x, y) = \frac{(2\mu_x\mu_y + C_1)(2\sigma_{xy} + C_2)}{(\mu_x^2 + \mu_y^2 + C_1)(\sigma_x^2 + \sigma_y^2 + C_2)}, \qquad (5)$$

where $\mu_x$ is the mean of x, $\sigma_x$ is the variance of x, $\sigma_{xy}$ is the covariance of x and y, and $C_1, C_2$ are constants. The value range of SSIM is [0, 1]. The larger the value of SSIM, the higher the structural similarity.

### 3.1 Using four pinhole images from single LEH

In this case, the self-radiation intensity distribution in the hohlraum was reconstructed using four pinhole images from single LEH. Fig. 3(a) shows the original 3D self-radiation intensity distribution used for the simulation (the upper LEH is not shown), Fig. 3(b) shows the reconstructed self-radiation distribution through the conventional SART method without spatial constraints (SART_wc), and Fig. 3(c) demonstrates the reconstructed self-radiation distribution by the proposed method. The three rings on the hohlraum wall are light spots formed by the laser injected from the upper and lower LEHs.

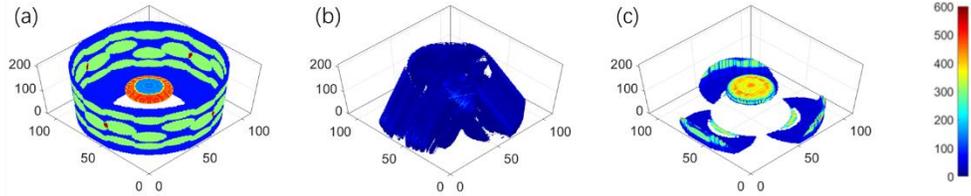

fig. 3. Comparison of the original and reconstructed 3D self-radiation intensity distribution: (A) original 3D self-radiation intensity distribution (except for the upper LEH), (B) reconstructed 3D self-radiation intensity distribution using SART_wc method, and (C) reconstructed 3D self-radiation intensity distribution using the proposed method.

The self-radiation intensity distribution reconstructed by the SART_wc method was approximately uniformly distributed and stretched in the area that should be zero. The proposed method can estimate the central shape of the plasma core. The radiation distribution of the hohlraum wall and LEH can also be partially observed.

The NMSEs between the simulated and reconstructed self-radiation intensity distribution were calculated for the SART_wc method and the proposed method: 0.9129 and 0.7443, respectively. Furthermore, the NMSEs of the central self-radiation area of 50 × 50 × 50 voxels were 0.7531 and 0.4401, respectively. The results suggested that the proposed method improved the estimation of the self-radiation distribution in the center of the hohlraum.

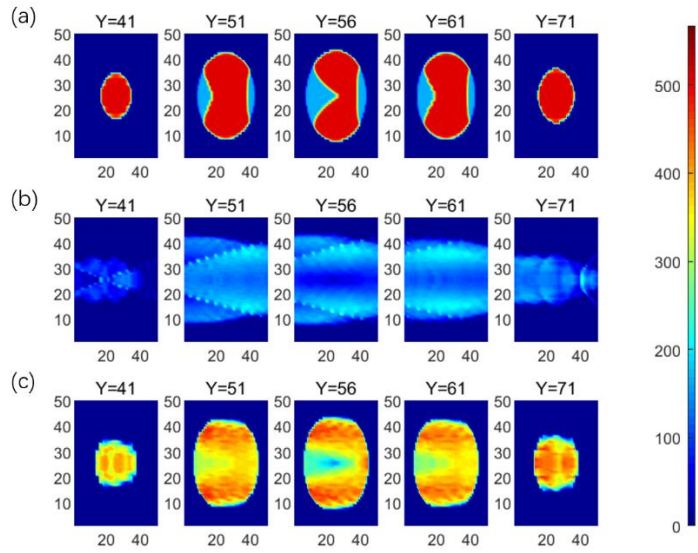

Fig. 4. Five sections of the central self-radiation area along the Y direction: (a) five sections of the simulated central self-radiation area, (b) sections reconstructed by the SART without spatial constraints, and (c) sections reconstructed by the proposed method.

Fig. 4 shows the comparison of five sections in the central self-radiation area along the Y direction between the conventional SART method and the proposed method. The reconstructed central self-radiation area by the SART_wc method became narrow due to the small projection angle. The general shape and distribution of the original plasma core were maintained using the proposed method. The NMSEs of different sections of the central self-radiation area along the X, Y, and Z directions were calculated to evaluate the reconstruction error inside the plasma core, as shown in Fig. 5. The proposed method was superior to the conventional algorithm in the X and Y directions. In the Z direction, the proposed method far exceeded the SART_wc method near the plasma core. The reason why NMSEs of the edge portion in the plasma core grew in size may be because the active area near the edge of the plasma core became smaller. In addition, the estimation for the central self-radiation area using the proposed method in the Z direction was slightly larger than the actual one because of the limitation associated with the projection angle.

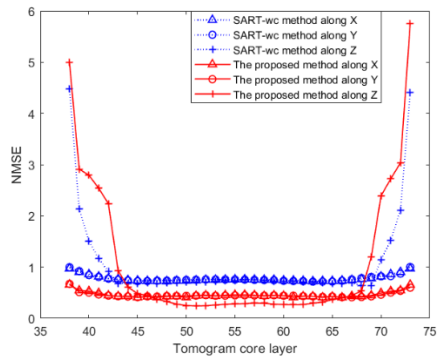

Fig. 5. Normalized mean square errors (NMSEs) of different sections in the central self-radiation area along the X, Y, and Z directions. The blue dotted lines represent NMSEs of the SART_wc method, and the red lines represent the proposed method.

The SSIMs between the simulated and reconstructed self-radiation intensity distribution were calculated for the SART_wc method and the proposed method: 0.1825 and 0.6891, respectively. Furthermore, the SSIMs of the central self-radiation area of $50 \times 50 \times 50$ voxels were 0.2534 and 0.5566, respectively. The SSIM of the proposed method significantly improved and exceeded 0.5, suggesting that a higher structural similarity was achieved. In addition, the SSIMs of different sections of the central self-radiation area along the X, Y, and Z directions were calculated, as shown in Fig. 6. The proposed method proved to be superior to the SART_wc method in all three directions, especially in the Z direction near the center of the hohlraum. The significant increase of the SSIM in the central self-radiation area when using the proposed method indicated that the proposed method achieved better estimation of the self-radiation area when compared with that achieved using the SART_wc method.

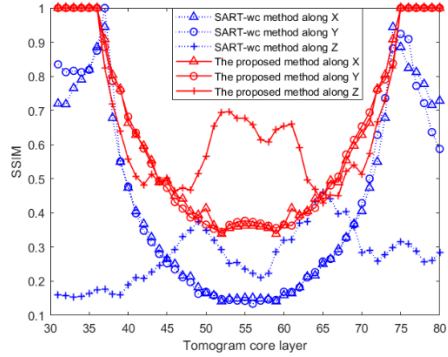

Fig. 6. SSIMs of different sections in the central self-radiation area along the X, Y, and Z directions. The blue dotted lines represent the SSIMs of the SART_wc method, and the red lines represent the proposed method.

The NMSEs between the pinhole images projected by the simulated self-radiation intensity distribution and the pinhole images projected by the reconstructed self-radiation intensity distribution were computed by two methods and are shown in Table 1:

Table 1: Normalized mean square errors (NMSE) of four pinhole images using two methods

| Methods | NMSE | | | |
|---|---|---|---|---|
| SART_wc method | 0.05172 | 0.02370 | 0.02350 | 0.02361 |
| Proposed method | 0.006864 | 0.0077864 | 0.008635 | 0.0082319 |

The results indicate that the NMSEs of two methods are less than 0.1 and that the reconstructed pinhole images of the proposed method exhibit higher qualities. Two original pinhole images and the pinhole images reconstructed using the proposed method are also shown in Fig. 7. The NMSEs between the original and reconstructed pinhole images using the proposed method are so small that no difference can be directly observed.

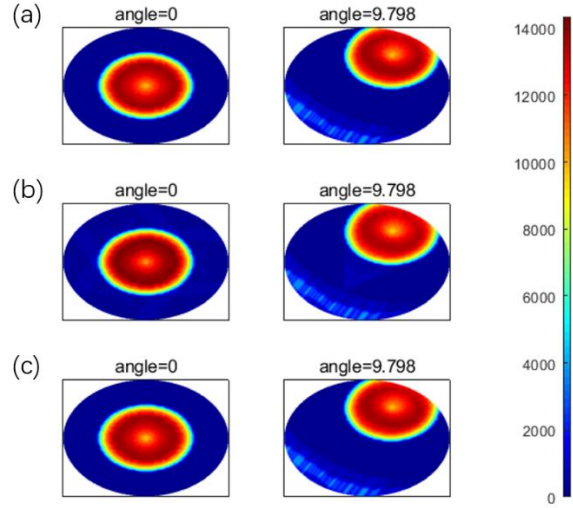

Fig. 7. Comparison between simulated and reconstructed pinhole images: (a) original pinhole images, (b) reconstructed pinhole images using the SART_wc method and (c) reconstructed pinhole images using the proposed method.

*3.2 Using eight pinhole images from a LEH and a diagnostic hole*

In this case, the self-radiation intensity distribution is reconstructed using eight pinhole images from a LEH and a diagnostic hole (two diagnostic holes). Fig. 8(a) shows the original 3D self-radiation intensity distribution used for the simulation (the upper LEH is not shown), Fig. 8(b) shows the reconstructed self-radiation distribution using single LEH and Fig. 8(c) demonstrates the reconstructed self-radiation distribution using two diagnostic holes.

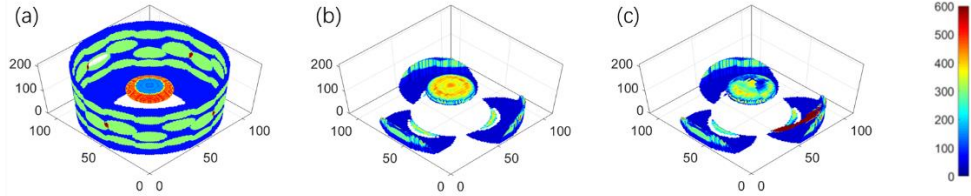

Fig. 8. Comparison of the original and reconstructed 3D self-radiation intensity distribution: (A) original 3D self-radiation intensity distribution (except for the upper LEH), (B) reconstructed 3D self-radiation intensity distribution using single LEH, and (C) reconstructed 3D self-radiation intensity distribution using two diagnostic holes.

Similar to the single LEH case, the proposed method can also estimate the central shape of the plasma core from two diagnostic holes. The NMSE between the simulated and reconstructed self-radiation intensity distribution was calculated for the proposed method using two diagnostic holes: 0.7452. Furthermore, the NMSE of the central self-radiation area of $50 \times 50 \times 50$ voxels was 0.3079. Compared with the single LEH case, the results suggested that using two diagnostic holes can improve the estimation of the self-radiation distribution in the center of the hohlraum.

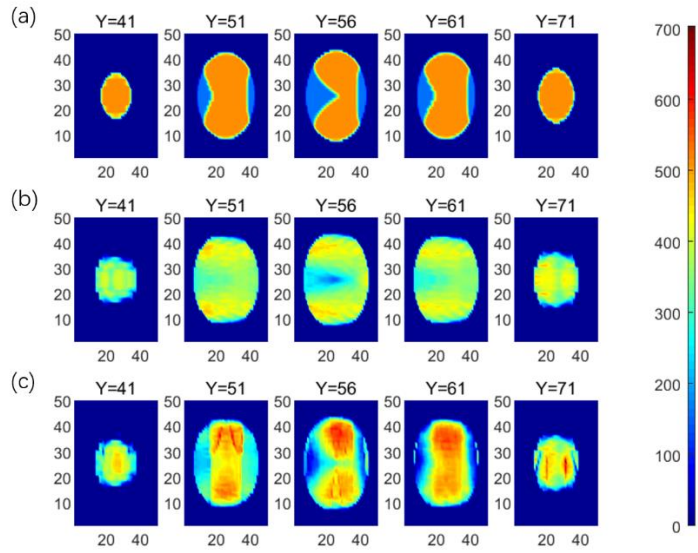

Fig. 9. Five sections of the central self-radiation area along the Y direction: (a) five sections of the simulated central self-radiation area, (b) sections reconstructed by single LEH, and (c) sections reconstructed by two diagnostic holes.

Fig. 9 shows the comparison of five sections in the central self-radiation area along the Y direction between single LEH and two diagnostic holes. The general shape and distribution of the original central self-radiation area were maintained better using two diagnostic holes. The NMSEs of different sections of the central self-radiation area along the X, Y, and Z directions were calculated to evaluate the reconstruction error inside the plasma core, as shown in Fig. 10. The NMSEs using two diagnostic holes were superior to that of single LEH in all three directions.

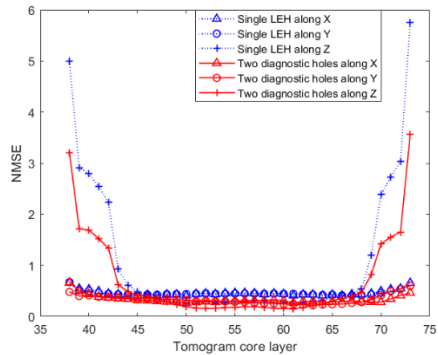

Fig. 10. Normalized mean square errors (NMSEs) of different sections in the central self-radiation area along the X, Y, and Z directions. The blue dotted lines represent NMSEs of using single LEH, and the red lines represent NMSEs of using two diagnostic holes.

The SSIM using two diagnostic holes between the simulated and reconstructed self-radiation intensity distribution was 0.6927. Furthermore, the SSIMs of the central self-radiation area of $50 \times 50 \times 50$ voxels was 0.5947. The SSIMs using two diagnostic holes improved both in the total and the central self-radiation area, suggesting that a higher structural similarity was achieved. In addition, the SSIMs of different sections of the central self-radiation area along the X, Y, and Z directions were calculated, as shown in Fig. 11. Using two diagnostic holes proved to have more advantages than using a single LEH in X and Y directions, especially in the area near the center of the hohlraum. The increase of the SSIM in the central self-radiation area indicated that using two diagnostic holes achieved better estimation of the plasma core compared with that achieved using a single LEH.

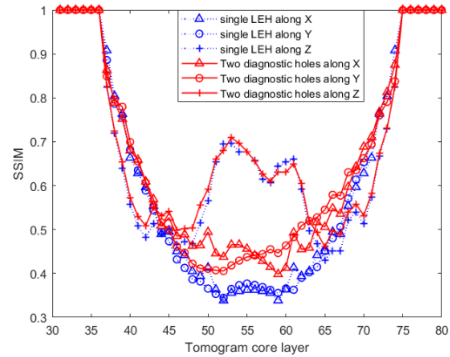

Fig. 11. SSIMs of different sections in the central self-radiation area along the X, Y, and Z directions. The blue dotted lines represent the SSIMs of using single LEH, and the red lines represent the SSIMs of using two diagnostic holes.

## 4. Conclusions

Generally, with four pinhole images at a maximum projection angle of 10°, the simulation results suggest that the proposed method can attain the approximate self-radiation distribution of the compressed plasma core in the center of the vacuum hohlraum. Compared with the NMSE between the simulated self-radiation intensity distribution and the reconstructed self-radiation intensity distribution, the proposed method achieves a lower NMSE than the SART method without spatial constraints, especially in the self-radiation area in the center of the vacuum hohlraum, which was 0.4401. The SSIM of the self-radiation intensity distribution reconstructed by the proposed method was more than 0.5. Compared with the simulated pinhole images and the reconstructed pinhole images, the NMSE of the proposed method was much lower than that of the conventional SART method without spatial constraints. Meanwhile, using more diagnostic holes could achieve better structural similarity and lower reconstruction error. Future efforts should focus on improving the accuracy of the estimated shape of the central self-radiation area to improve the accuracy of the reconstructed 3D distribution. Furthermore, the X-ray framing camera can be added to obtain time-resolved 3D self-radiation images.

## Funding

National Natural Science Foundation of China (NSFC) (11675014, 11675157, and 11805180).

## Acknowledgments

Shijian Li and Qiangqiang Wang contributed equally to this work. The authors would like to thank Prof. Molin Ge for the valuable discussions.

## Disclosures

The authors declare no conflicts of interest.

## References

1. J. D. Lindl, P. Amendt, R. L. Berger, S. G. Glendinning, S. H. Glenzer, S. W. Haan, R. L. Kauffman, O. L. Landen, and L. J. Suter, "The physics basis for ignition using indirect-drive targets on the national ignition facility," Phys. Plasmas **11**, 339-491(2004).
2. J. A. Koch, O. L. Landen, T. W. Barbee, P. Celliers, L. B. Da Silva, S. G. Glendinning, B.A. Hammel, D. H. Kalantar, C. Brown, J. Seely, G. R. Bennett, and W. Hsing, "High-energy x-ray microscopy techniques for laser-fusion plasma research at the National Ignition Facility," Appl. Opt. **37**, 1784-1795 (1998).
3. F. J. Marshall, J. A. Delettrez, R. Epstein, V. Yu. Glebov, D. R. Harding, P. W. McKenty, D. D. Meyerhofer, P. B. Radha, W. Seka, S. Skupsky, V. A. Smalyuk, J. M. Soures, C. Stoeckl, R. P. J. Town, and B. Yaakobi, "Direct-drive high-convergence-ratio implosion studies on the OMEGA laser system," Phys. Plasmas **7**, 2108 (2000).
4. R. E. Turner, "FY'03 OMEGA Summary for LLE Annual Report," Lawrence Livermore National Laboratory Report No. UCRL-TR-155665 (2003).
5. K. Ren, S. Liu, L. Hou, H. Du, G. Ren, W. Huo, L. Jing, Y. Zhao, Z. Yang, M. Wei, K. Deng, L. Yao, Z. Li, D. Yang, C. Zhang, J. Yan, G. Yang, S. Li, S. Jiang, Y. Ding, J. Liu, and K. Lan, "Direct measurement of x-ray flux for a pre-specified highly-resolved region in hohlraum," Opt. Express **23**, A1072-A1080 (2015).
6. K. Ren, S. Liu, X. Xie, H. Du, L. Hou, L. Jing, D. Yang, Y. Zhao, J. Yan, Z. Yang, Z. Li, J. Dong, G. Yang, S. Li, Z. Cao, K. Lan, W. Huo, J. Liu, G. Ren, Y. Ding and S. Jiang, "First exploration of radiation temperatures of the laser spot, re-emitting wall and entire hohlraum drive source," Sci. Rep. **9**, 5050 (2019).
7. L. Gao, J. Liang, C. Li and L. V. Wang, "Single-shot compressed ultrafast photography at one hundred billion frames per second," Nature **516**, 74–77 (2014).
8. J. Liang, L. Zhu, and L.V. Wang, "Single-shot real-time femtosecond imaging of temporal focusing," Light Sci. Appl. **7**, 42 (2018).
9. D. S. Clarka, C. R. Weber, J. L. Milovich, J. D. Salmonson, A. L. Kritcher, S. W. Haan, B. A. Hammel, D. E. Hinkel, O. A. Hurricane, O. S. Jones, M. M. Marinak, P. K. Patel, H. F. Robey, S. M. Sepke, and M. J. Edwards, "Three-dimensional simulations of low foot and high foot implosion experiments on the National Ignition Facility," Phys. Plasmas **23**, 056302 (2016).
10. D.S. Clark, C.R. Weber, A.L. Kritcher, J.L. Milovich, P.K. Patel, S.W. Haan, B.A. Hammel, J.M. Koning, M.M. Marinak, M.V. Patel, C.R. Schroeder, S.M. Sepke and M.J. Edwards, "Modeling and projecting implosion performance for the National Ignition Facility," Nucl. Fusion **59**(3), 032008.1-032008.9. (2019).
11. Y. Pu, T. Huang, F. Ge, S. Zou, F. Wang, J. Yang, S. Jiang and Y. Ding, "First integrated implosion experiments on the SG-III laser facility," Plasma Phys. Control. Fusion **60**, 085017(2018).
12. G. N. Minerbo, J. G. Sanderson, D. B. van Hulsteyn, and P. Lee, "Three-dimensional reconstruction of the x-ray emission in laser imploded targets," Appl. Opt. **19**, 1723–1728 (1980).
13. D. Ress, R. A. Lerche, L. Da Silva, "Demonstration of an x ray ring-aperture microscope for inertial-confinement fusion experiments," Appl. Phys. Lett. **60**, 410–412 (1992).
14. G. Gillman and I. Macleod, "Reconstruction of x-ray sources from penumbral images," Comput. Graph. Image Process. **11**, 227–241 (1979).
15. T. Ueda, S. Fujioka, S. Nozaki, R. Azuma, Y.-W. Chen, and H. Nishimura, "A uniformly redundant imaging array of penumbral apertures coupled with a heuristic reconstruction for hard x-ray and neutron imaging," Rev. Sci. Instrum. **81**, 073505 (2010).
16. J. Arai, F. Okano, H. Hoshino, and I. Yuyama, "Gradient-index lens-array method based on real-time integral photography for three-dimensional images," Appl. Opt. **37**, 2034–2045 (1998).
17. X. Wang and H. Hua, "Theoretical analysis for integral imaging performance based on micro scanning of a microlens array," Opt. Lett. **33**, 449–451 (2008).
18. L. Zhang, Y. Yang, X. Zhao, Z. Fang, and X. Yuan, "Enhancement of depth-of-field in a direct projection-type integral imaging system by a negative lens array," Opt. Express **20**, 26021–26026 (2012).
19. X. Xiao, B. Javidi, M. Martinez-Corral, and A. Stern, "Advances in three-dimensional integral imaging: sensing, display, and applications," Appl. Opt. **52**, 546–560 (2013).
20. H. Deng, Q.-H. Wang, F. Wu, Ch.-G. Luo, and Y. Liu, "Cross-talk-free integral imaging three-dimensional display based on a pyramid pinhole array," Photon. Res. **3**, 173-176 (2015).
21. Y.-W. Chen, T. Kohatsu, S. Nozaki, and R. Kodama, "Heuristic reconstruction of three-dimensional laser-imploded targets from x-ray pinhole images," Rev. Sci. Instrum. **74**, 2236–2239 (2003).
22. P. Xu, Y. Bai, X. Bai, B. Liu, X. Ouyang, B. Wang, W. Yang, Y. Gou, B. Zhu, and J. Qin, "Three-dimensional reconstruction of laser-imploded targets from simulated pinhole images," Appl. Opt. **51**, 7820-7825 (2012).
23. A. H. Andersen, and A. C. Kak. "Simultaneous Algebraic Reconstruction Technique (SART): A superior implementation of the ART algorithm," Ultrasonic Imaging **6**(1),81-94 (1984).
24. K. Mueller and R. Yagel, "Rapid 3-D cone-beam reconstruction with the simultaneous algebraic reconstruction technique (SART) using 2-D texture mapping hardware," IEEE T. Med. Imaging **19**(12), 1227-1237 (2000).
25. H. Lee, B. Song, J. Kim, J. J. Jung, H. Li, S. Mutic, J. C. Park, "Variable step size methods for solving simultaneous algebraic reconstruction technique (SART)-type CBCT reconstructions," Oncotarget **8**,33827-33835 (2017).
26. S. Zhang, G. Geng, G. Cao, Y. Zhang, B. Liu and X. Dong, "Fast Projection Algorithm for LIM-Based Simultaneous Algebraic Reconstruction Technique and Its Parallel Implementation on GPU," IEEE Access **6**, 23007-23018 (2018).

27. T. Hobiger, T. Kondo, and Y. Koyama, "Constrained simultaneous algebraic reconstruction technique (C-SART) —a new and simple algorithm applied to ionospheric tomography," Earth Planet Sp. **60**, 727–735 (2008).
28. Y.M., Levakhina, J. Müller, R.L. Duschka, F. Vogt, J. Barkhausen and T.M. Buzug, "Weighted simultaneous algebraic reconstruction technique for tomosynthesis imaging of objects with high-attenuation features," Med. Phys. **40**,031106 (2013).
29. D. Ji, G. Qu, and B. Liu, "Simultaneous algebraic reconstruction technique based on guided image filtering," Opt. Express **24**, 15897-15911 (2016).
30. A. Boudjelal, Z. Messali, A. Elmoataz, and B. Attallah, "Improved Simultaneous Algebraic Reconstruction Technique Algorithm for Positron-Emission Tomography Image Reconstruction via Minimizing the Fast Total Variation," Journal of Medical Imaging and Radiation Sciences **48**(4),385-393(2017).
31. Z. Wang, A. C. Bovik, H. R. Sheikh, and E. P. Simoncelli, "Image Quality Assessment: From Error Visibility to Structural Similarity," IEEE T. Image Processing **13**(4),600-612(2004).